\documentstyle[prb,aps,twocolumn,epsfig]{revtex}
\newcommand\ggg{Gd$_3$Ga$_5$O$_{12}$}
\newcommand\me{O.~A.~Petrenko}
\newcommand\MC{Monte Carlo}
\newcommand\MCS{Monte Carlo simulation}
\newcommand\Kag{{\it Kagom\'{e}}}
\newcommand\pyr{pyrochlore}

\newcommand\afm{anti\-ferro\-magnet}
\newcommand\afmc{anti\-ferro\-magnetic}
\newcommand\PRL[3]{Phys. Rev. Lett. {\bf {#1}}, {#2} ({#3})}
\newcommand\PRB[3]{Phys. Rev. B {\bf {#1}}, {#2} ({#3})}
\newcommand\JPCM[3]{J. Phys.: Condens. Matter {\bf {#1}}, {#2} ({#3})}
\newcommand\PhysB[3]{Physica B {\bf {#1}}, {#2} ({#3})}
\newcommand\JPSJ[3]{J. Phys. Soc. Japan {\bf {#1}}, {#2} ({#3})}
\begin{document}
\draft
\twocolumn[\hsize\textwidth\columnwidth\hsize\csname @twocolumnfalse\endcsname
\title{Classical Heisenberg \afm\ on a garnet lattice: a \MCS}
\author{\me$^{1,2}$ and D.~McK~Paul$^2$}
\address{$^1$ ISIS Facility, Rutherford Appleton Laboratory, Chilton, Didcot, OX11~0QX, UK}
\address{$^2$ University of Warwick, Department of Physics, Coventry, CV4~7AL, UK}
\date{\today}
\maketitle
\begin{abstract}
We have studied a classical \afm\ on a garnet lattice by means of \MCS s in an attempt
to examine the role of geometrical frustration in Gadolinium Gallium Garnet, \ggg\ (GGG).
Low-temperature specific heat, magnetisation, susceptibility, the autocorrelation function $A(t)$
and the neutron scattering function $S(Q)$ have been calculated for several models including different
types of magnetic interactions and with the presence of an external magnetic field applied along the
principal symmetry axes. A model, which includes only nearest-neighbour exchange, $J_1$, neither orders
down to the lowest temperature nor does it show any tendency towards forming a short-range coplanar
spin structure. This model, however, does demonstrate a magnetic field induced ordering below $T\sim 0.01J_1$.
In order to reproduce the experimentally observed properties of GGG, the simulated model must include
nearest neighbour exchange interactions and also dipolar forces. The presence of weak next-to-nearest
exchange interactions is found to be insignificant. In zero field $S(Q)$ exhibits diffuse magnetic
scattering around positions in reciprocal space where \afmc\ Bragg peaks appear in an applied magnetic field.
\end{abstract}
]

\flushbottom
\parskip 0mm
\noindent
\section{Introduction}

The introduction of frustration to magnetic systems leads to extra degeneracy for the ground
state in addition to the degeneracy resulting from the symmetry of magnetic Hamiltonian.
The larger this additional degeneracy, the more likely frustration is to cause dramatic changes
in the magnetic properties of the system, such as the absence of long range order even at the
lowest temperature. Geometrical frustration has been one of the key issues in magnetism
for at least twenty years. A recent wave of theoretical papers \cite{Pyro_theor,Kagome_theor}
as well as publications dealing with \MCS\ \cite{Pyro_MC,Siddharthan_99,Kagome_MC,Chalker_92} 
has emphasised the unusual magnetic properties of geometrically frustrated systems. The question
of whether the frustration leads to a disordered gapped state or to long-range Ne\'{e}l type order
in different types of geometry is still under debate. Current efforts seem to be concentrated
around two types of lattices: the \Kag\ lattice \cite{Kagome_theor,Kagome_MC,Triang_MC} and the \pyr\
lattice \cite{Pyro_theor,Pyro_MC}. Recently has it been established that the \pyr\ lattice
represents the only simple system for which the additional degree of freedom caused by the
frustration is extensive -- it is proportional to the number of spins involved \cite{Moessner_98}.

The growth of theoretical interest in the \pyr\ lattice, a lattice of corner-sharing tetrahedra, is driven largely
by experimental discoveries \cite{Pyro_exp}. There are many chemically clean \pyr s (some of which  may
be produced as single crystals \cite{Bala_98}) with different types of magnetic atoms and interactions,
which allows one to pick the most suitable one for study and for comparison with a particular theoretical
model. By studying the phenomenon in general a much better understanding of the magnetic properties of
individual compounds can be achieved. The same reasoning applies to another geometrically frustrated
system - an \afm\ on a \Kag\ lattice, where SrCr$_{9p}$Ga$_{12-9p}$O$_{19}$ \cite{SrCrGaO},
jarosites \cite{jarosites} and some other compounds \cite{other_Kagome} provide quite a variety
of model systems.  

\begin{figure}
\centerline{\centerline{\psfig{figure=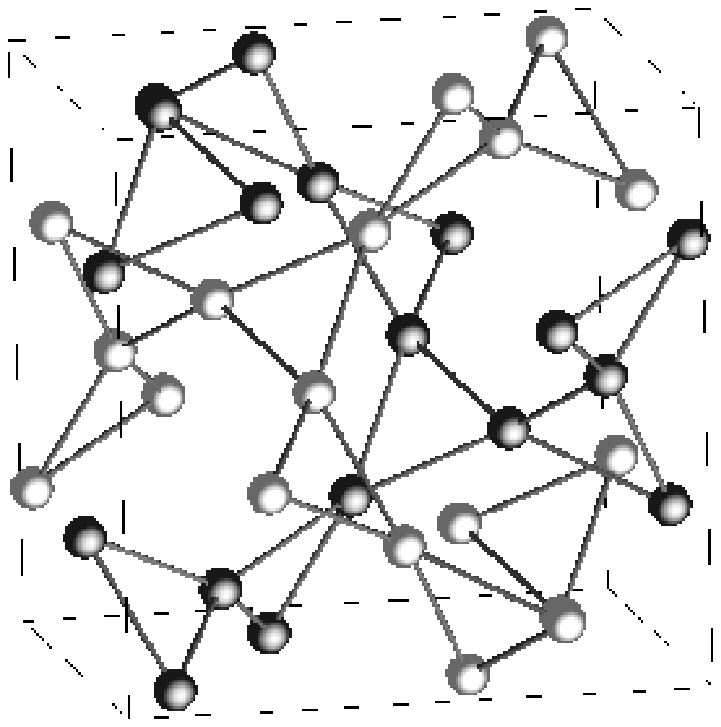,width=\columnwidth}}}
\caption{Positions  of the magnetic Gd ions in a garnet  structure. There are 24 magnetic ions per
unit cell,  they are divided into two interpenetrating sublattices.}
\label{CrystStr}
\end{figure}

Gadolinium gallium garnet, \ggg , is a {\it unique} example of an \afm\ on the garnet
lattice. There are no other compounds matching its magnetic properties. In GGG (space group
Ia$\bar{3}$d) the magnetic Gd ions are located on two interpenetrating, corner-sharing triangular
sublattices, where the triangles of spins do not lie in the same plane -- the angle between two
nearest triangles is equal to the angle between the diagonals of a cube, 70.5$^\circ$ (see
Fig. \ref{CrystStr}). In this compound the triangular arrangement of the nearest spins is combined
with complete exchange isotropy (the single-ion anisotropy is negligibly small \cite{GGG_isotrop})
and with a relatively strong dipole-dipole energy. Although the magnetic properties of various
garnets have been thoroughly studied during the past half century, the analogy between any of them
and GGG is not straightforward. All other magnetic garnets order at some low temperature, while GGG
does not. No long range magnetic order has been detected in GGG down to 25~mK \cite{Ramirez_91}, 
while other gallium garnets based on Dy, Nd, Sm and Er, rather than Gd, have been found to be 
magnetically ordered at temperatures below 1~K \cite{Filippi_Onn}. The nearest analogy to GGG would
probably be found among the  Mn-based garnets, where the single-ion anisotropy is also very small.
However  two similar magnetically isotropic garnets, Mn$_3$Al$_2$Ge$_3$O$_{12}$ and
Mn$_3$Al$_2$Si$_3$O$_{12}$ \cite{MnAlgarnets}, also order. Most likely this is due to the
presence of relatively strong next-to-nearest exchange interactions. If and when the degeneracy of
the ground state is removed and the system undergoes a phase transition to a long-range ordered state,
almost all complications disappear. The magnetic ground state and the main interactions are known
from experiment and theoretical calculations are straightforward. Numerical estimates exist to at
least the accuracy that experiments currently attain. However a theoretical model describing
adequately the magnetic properties of GGG still has to be developed.

This paper presents the results of classical \MC\ simulations of the magnetic properties of the Heisenberg
antiferromagnet on a garnet lattice. While some of the initial results related to the GGG have been briefly
reported in our neutron scattering  papers \cite{Petrenko_98,Petrenko_99}, where they have been used to
explain the obtained experimental data and also to predict possible experiments, here we take a more general
approach to the problem. We address issues which are a not necessarily directly related to GGG, but are
interesting from a theoretical point of view, {\it e.g.} we discuss properties of a model which includes
nearest-neighbour exchange interactions only. Where possible we compare with the results of simulations for
the \pyr\ and \Kag\ lattices and show, that an \afm\ on a garnet lattice is yet another highly frustrated magnetic
system exhibiting a number of unusual and intriguing properties.

\section{Simulation Models}

We consider the Hamiltonian
\begin{equation}
\hat{\cal H}=\sum_{<i,j>}J_{ij}\;{\bf S_i S_j}\; +
   D\sum_{<i,j>}\left [\frac{\bf S_i S_j}{r_{ij}^3} - 3\frac{({\bf S_i r_{ij}})({\bf S_j  r_{ij}})}{r_{ij}^5}\right ],
\end{equation}
where the spins ${\bf S}_i$ are classical, three-component vectors on the Gd$^{3+}$ sites of a garnet
lattice, S=7/2 as in GGG. The first term is the exchange interaction, the second term is the 
dipole-dipole interaction between the magnetic moments.

The original idea to simulate the magnetic properties of GGG using MC methods belongs to Kinney 
and Wolf \cite{Kinney_79}, who calculated the temperature dependence of the specific heat and by 
comparing the results with the experimental data have obtained the amplitudes of the nearest and
next to nearest neighbour exchange interactions $J_1$, $J_2$ and $J_3$. More recently Schiffer
{\it et.~al} calculated the magnetic phase boundary and have investigated the magnetic structure
of GGG in an applied field \cite{Schiffer_94}. We use the same value of the nearest exchange constant
as Kinney and Wolf \cite{Kinney_79}, $J_1=0.107$~K, because it produces good estimates for
the temperature dependence of the susceptibility and also for the saturation field of the
magnetisation \cite{Fischer_73}. However, as will be shown later, the values of $J_2$ and $J_3$
quoted in Ref. \cite{Kinney_79} are not essential: as long as they are small in comparison with $J_1$,
they do not change significantly the predicted magnetic properties of GGG and therefore can not be reliably
determined from the MC simulations. 

The strength of the dipole-dipole interaction, $D$, is defined by the distances between the $i$th
and $j$th spins. In GGG the Gd$^{3+}$ sites are separated by $\frac{\sqrt{6}}{8}a = 3.781$~\AA ,
where $a=12.349$~\AA\ is the lattice constant at low temperature. Therefore we use $D_{dd}=0.0457$~K for
the strength of the nearest-neighbour dipolar interaction. It is very important and at the same time
very difficult to simulate reliably such a long range interaction as the dipole-dipole one. For some
simpler lattices, for example, a 2D-square lattice \cite{MacIsaac_98}, or for highly anisotropical systems,
such as {\it spin ice} pyrochlores \cite{Hertog_00}, the Ewald summation technique can be used to treat
the long-range nature of the dipole-dipole interaction. In case of Heisenberg spins  located on the complicated
lattice of GGG, however, there is no option but to introduce a cut-off range, $R_0$, and to neglect the
dipole-dipole interaction for all distances larger than $R_0$. Previous simulations have restricted the
dipole-dipole interaction to a third neighbour \cite{Kinney_79,Schiffer_94}, while in our model $R_0$
has been extended to include the fourth neighbour. We have also made several test runs to compare
the simulation results for this model  with both shorter (to a third neighbour) and also longer (ten
neighbours) cut-off ranges and have found no significant difference, which suggests that this model describes
the dipolar force reasonably well.
The dipolar interaction between two magnetic moments decays as $\frac{1}{R^3}$, the number of neighbours
in a shell $\delta R$ is proportional to $R^2$, therefore the dipole-dipole energy should decay 
only relatively slowly, as $\frac{1}{R}$. In reality, however, the extension of the cut-off range
from a third neighbour to a fourth one does not change significantly either the total dipolar
energy, nor the overall system energy. A possible answer to this puzzle might be related to the fact
that all magnetic interactions in GGG, including the dipolar one, are frustrated: the contribution of
the individual magnetic moments to the total system energy is mutually cancelled or nearly cancelled,
therefore for each magnetic moment only the local surroundings influence the choice of magnetic
orientation. Similar observations have been made during recent \MCS s on \pyr\ lattice which
included long-ranged dipole-dipole as well as short-ranged exchange interactions \cite{Siddharthan_99}.

MC simulations have been performed for lattice sizes $L\!\times\!L\!\times\!L$, with $L\!=\!3$ to 9
unit cells, containing 648 to 17496 spins. Significantly larger lattice sizes, than previously used,
have ensured that the magnetic correlation length in the disordered phase does not exceed the system size.
Simulations with larger lattice sizes have improved the resolution of the calculated scattering
function, $S(Q)$, in an applied magnetic field allowing us to resolve clearly individual magnetic Bragg peaks.
A standard Metropolis algorithm with periodic or open boundary condition has been employed; up to several
millions \MC\ steps per spin (MCS) were performed at the lowest temperatures. Where possible an
attenuation factor $\delta S$ has been introduced in such a way that roughly 50\% of the attempted spin
moves were accepted \cite{Reimers_92}, which has resulted in a dramatic increase of the spin relaxation
rate. For the simulations in a magnetic field this procedure has been abandoned to permit the system
to make abrupt structural changes.

The magnetic field is assumed to be applied along the $(100)$ direction unless otherwise stated.

\section{Results and discussion}

\subsection{Zero external field properties}

We begin by addressing the issue of the phase transition at low temperature in zero magnetic field.
In GGG no sign of long range magnetic  order has been found down to 25~mK \cite{Ramirez_91}, moreover,
frustration induced spin freezing has been suggested at temperatures  below  125~-~135~mK on the basis
of single crystal magnetisation measurements: the susceptibility is frequency dependent, and the static
magnetisation is different for field cooling and zero field cooling \cite{Schiffer_95}. However, neutron
scattering experiments show that at the lowest temperatures the magnetic system is not frozen
completely \cite{Petrenko_98}. It rather behaves as a mixture of a liquid and solid states.  

\begin{figure}
\centerline{\psfig{figure=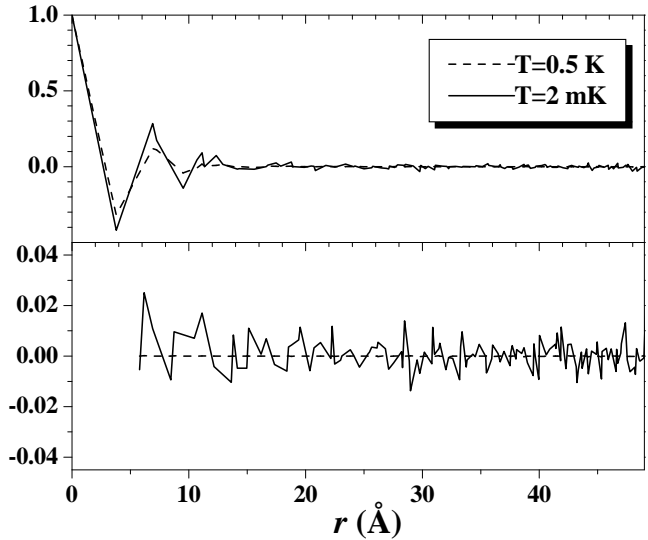,width=\columnwidth}}
\caption{Correlation function for a system of $5\!\times\!5\!\times\!5$ unit cell sizes (3000 spins), which
includes only nearest neighbour exchange interaction, $J_1$. Top picture shows correlation between
spins belonging to the same sublattice, bottom -- correlation between sublattices.}
\label{CorrF}
\end{figure}

\begin{figure}
\centerline{\psfig{figure=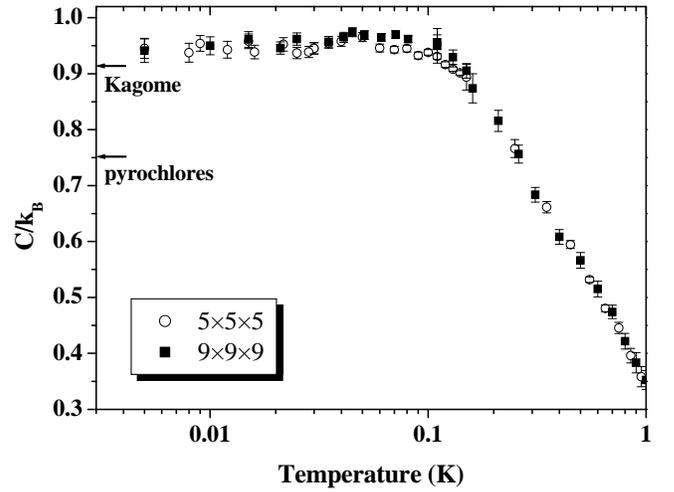,width=\columnwidth}}
\caption{Specific heat temperature dependence for an open boundary conditions model of 
$5\!\times\!5\!\times\!5$ and $9\!\times\!9\!\times\!9$ sizes which includes only the nearest 
neighbour exchange interaction, $J_1$. Up to $4\times10^6$ MCS have been performed at 
lower temperatures.}
\label{Cp}
\end{figure}

The first thing to notice is that the simulation model, which includes only nearest-neighbour exchange,
$J_1$, does not show any sign of a phase transition down to at least $T\!=\!1$~mK (which is less than
0.1\% of the exchange energy $JS^2$). Several measured quantities show that the system remains in a
spin-liquid (or, following Villain \cite{Villain_79}, a cooperative paramagnet) phase: averaging over
sufficiently long time gives zero magnetic moment on each site, the scattering function $S(Q)$ does not
show any sharp peaks, the magnetic correlation length $Q(r)\!\equiv <\!{\bf S}(0)\!\cdot\!{\bf S}(r)\!>$
does not exceed the system size (see Fig. \ref{CorrF}). In fact, close inspection of Fig.~\ref{CorrF}
reveals that correlations are very small beyond the first unit cell. In addition there is no obvious
maximum or cusp in the heat capacity temperature dependence (see Fig. \ref{Cp}). To test the suspicion
that at low temperature \MCS s are not effective enough in allowing the system to reach equilibrium,
we have checked whether the simulation results depend upon the starting conditions. No difference
in results have been noticed when starting calculations from an initially random or a 120$^{\circ}$-degree
planar triangular state.   

The low-temperature specific heat itself is an important thermodynamic quantity, whose value is sensitive
to the presence of zero modes \cite{Chalker_92} and quartic modes \cite{Moessner_98}. In the \pyr\ lattice
each quadratic mode contributes $k_B/2$ to the heat capacity, each quartic mode $k_B/4$ and zero
modes do not contribute at all, thus reducing the zero-temperature specific heat to
3$k_B$/4 \cite{Moessner_98}, while in \Kag\ lattice it is reduced to 11$k_B$/12 \cite{Chalker_92}.
Our initial calculation on a relatively small system with periodic boundary conditions showed that
$C(\!T=\!0)$ was indistinguishable from unity within the accuracy of the simulations. However, prompted
by the comparison with the \Kag\ and \pyr\ lattice results, we have performed much longer \MC\ runs
on much bigger systems with open boundary conditions (we use open boundary conditions in order to avoid
the imposition of periodicity on a potentially incommensurate magnetic system). As can be seen from
Fig.~\ref{Cp}, $C(T\!=\!0)\!\approx\!0.94(2)$ with the accuracy of the calculation sufficiently high to claim that
it is actually below unity. There is no significant difference in $C(\!T=\!0)$ calculated for systems
of $5\!\times\!5\!\times\!5$ and $9\!\times\!9\!\times\!9$ containing 3000 and 17496 spins
respectively. 

\begin{figure}
\centerline{\psfig{figure=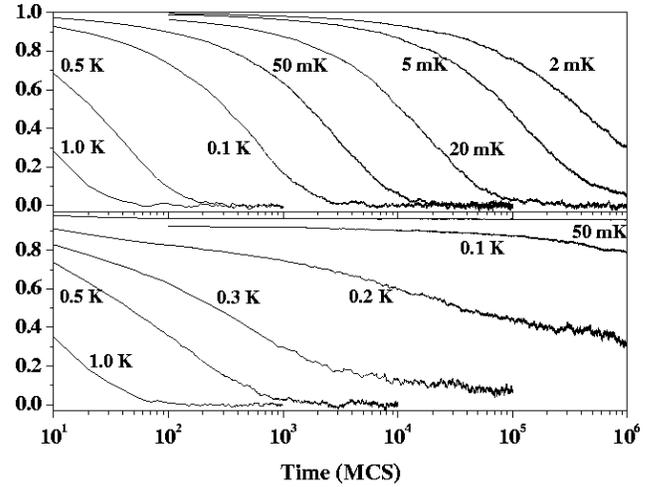,width=\columnwidth}}
\caption{Time dependence (in \MC\ steps per spin) of the autocorrelation function $\frac{1}{N}<S(0)S(t)>$ 
for various temperature from $T=1$~K down to 2~mK in a model which includes: a) only the nearest 
neighbour exchange interaction, $J_1$, b) the nearest neighbour exchange interaction and the dipole-dipole
interactions up to fourth neighbour.  System with a lattice size $9\!\times\!9\!\times\!9$ unit cells has 
been used for this calculations.}
\label{AutoCorr}
\end{figure}

The introduction of the dipolar interactions slows down the spin-relaxation process. Fig. \ref{AutoCorr}
displays the time dependence (time is measured in MCS) of the autocorrelation function
$A(t)=\frac{1}{N}\sum<\!{\bf S}_i(0){\bf S}_i(t)\!>$ for the two models: with (bottom) and without (top)
dipolar forces. The model which includes dipole-dipole interactions does not show noticeable relaxation
by $T=50$~mK, while the model with only nearest exchange interactions is still relaxing even at an order
of magnitude lower temperature.  The difference between the autocorrelation function for these two
models is evident at all temperatures below 0.5~K, which approximately coincides with the nearest
neighbour dipolar energy, $D_{dd} \times S^2$.

Dealing with very slow relaxing spin systems and a potential spin-glass transition it is essential to
ensure that the simulation time is longer than the equilibration time.  In practice the first $t_0$ 
MCS are used only for equilibration and then calculations and averaging are carried out during the next
$t_0$ steps. An estimation of an appropriate value of $t_0$ could be obtained following the procedure
introduced by Bhatt and Young \cite{Bhatt_85}, where the spin-glass susceptibility, $\chi_{SG}$, has
been calculated in two different ways.  In the first method we calculate an overlap between two
uncorrelated sets of spins which approaches $\chi_{SG}$ from below, if $t_0$ is shorter than the
equilibration time. In the second approach the four-spin-correlation function is calculated, which
approaches $\chi_{SG}$ from above, if $t_0$ is small. The $t_0$ is considered to be long enough
and the results are accepted only if the two estimates of $\chi_{SG}$ agreed. In a GGG model which
 includes both exchange and dipolar interactions $t_0$ becomes enormously long at low temperatures.
In fact even during the runs with $t_0=10^6$ MCS the results showed no agreement between the
 two approaches for all temperatures below $T=100$~mK.  Therefore the results of calculations in
zero field for a model which includes dipole-dipole interactions could not be considered as reliable
below this temperature. The problem of long equilibration times is removed by the application of an
external magnetic field.

The results of the simulations with the model, which takes into account only the nearest neighbour
exchange interaction, $J_1$, fits well the experimental neutron scattering function $S_p(Q)$ at
all temperatures above 140~mK \cite{Petrenko_98}. The neutron scattering function $S_p(Q)$ for a
powder sample is calculated as:
\begin{equation}
S_p(Q)=f(Q)^2\sum_{i,j}<{\bf S}_i{\bf S}_j>\frac{\sin(Qr)}{Qr},
\label{Sp}
\end{equation}
where $f(Q)$ is the magnetic form factor.  $S_p(Q)$ has several  broad diffuse scattering peaks (see
Fig.~5 in Ref. \cite{Petrenko_98}), whose intensity  increases as the temperature decreases in
agreement with the experiment. The introduction of the dipole-dipole interaction at these temperatures
does not change $S_p(Q)$ significantly.

\begin{figure}
\centerline{\psfig{figure=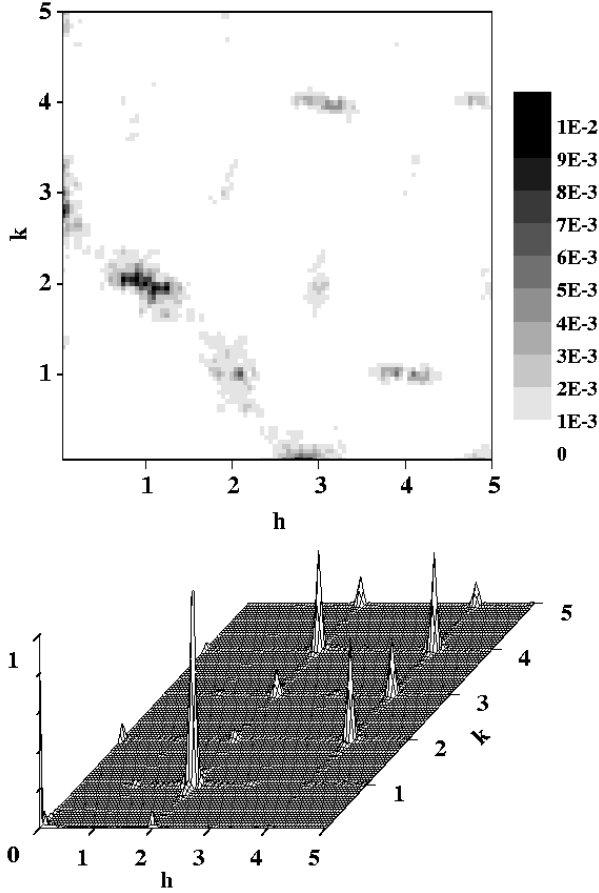,width=\columnwidth}}
\caption{Simulated single crystal neutron scattering  intensity of in the $(hk0)$ plane at $T=0.2$~K in a zero field (top panel) and in a field of $H=1.06$~T applied along $(001)$ direction (bottom panel). 
The data have been obtained according to formula~(\ref{Sxt}) for a model size of $9\!\times\!9\!\times\!9$ unit cells with open boundary conditions.}
\label{Sq}
\end{figure}

A somewhat unexpected results have been obtained earlier \cite{Petrenko_N2M} for a single crystal
neutron scattering function, calculated as:
\begin{equation}
S_{xt}(Q)=(f(Q)\sum_n^N{\bf q}_n e^{i{\bf Q}r_n})^2,
\label{Sxt}
\end{equation}
where ${\bf q}_n$ is the magnetic interaction vector.  Even at temperature well above $T=140$~mK, where 
there is no problem from very long equilibration times, $S_{xt}(Q)$ demonstrates a tendency to form
incommensurate peaks around integer positions in the reciprocal space (see Fig.~4 in 
Ref.~\cite{Petrenko_N2M}). The intensity of these incommensurate peaks is much lower than the expected
intensity of the true long-range order Bragg peaks, and their width is determined by the system size. 
The exact position of these peaks in reciprocal space is not fixed, it may change from one ``snap shot''
of $S_{xt}(Q)$ to another. Only after averaging significantly large amount of the ``snap shots'' (from
several dozens to several hundreds) a clear picture of the short-range incommensurate magnetic order was
obtained.  However, this is most likely to be an artificial effect caused by the periodic boundary
conditions: when they are removed, the effect of splitting seems to disappear. Top panel of Figure \ref{Sq}
shows simulated  single crystal neutron scattering  intensity of GGG in the $(hk0)$ plane at $T=0.2$~K.

Another interesting aspect of this study is to investigate how the ratio of exchange to dipolar
interactions influences properties of the Heisenberg \afm\ on a garnet lattice. In GGG $J_1$
is about twice the strength of $D_{dd}$ and there is no magnetic order, while in Mn-based
garnets~\cite{MnAlgarnets} the ratio $J_1/D_{dd}$ is slightly higher and they do order. For instance,
in Mn$_3$Al$_2$Ge$_3$O$_{12}$, which undergoes an \afmc\ phase transition to a
120$^\circ$-structure at $T_N\!=\!6.65$~K, the $J_1\!=\!0.57$~K \cite{Valyanskaya_76} is more than
ten times stronger than $D_{dd}$ in GGG. Our simulation shows that this fact alone could not lead to the
appearance of the long-range magnetic order. In a model, where $J_1$ has been increased up to a hundred
times keeping the $D_{dd}$ value fixed,  the ground state remained disordered. However, the introduction of
the next-to-nearest exchange interaction with a value cited  in \cite{Valyanskaya_76}, $J_2=0.12$~K, does
make a difference: the system immediately undergoes a phase transition to a LRO state, which reveals
itself clearly both as a cusp in a heat capacity temperature dependence and as peaks in the scattering
function, $S(Q)$.

\subsection{Magnetic properties in an applied field} 

As has been mentioned above, in an applied magnetic field the problem of long equilibration times
is much less severe, which gives us an excellent opportunity to investigate the magnetic phase diagram
of GGG in detail. A phase transition to a LRO state in magnetic field was detected by calculating the
specific heat temperature dependence in constant field or by calculating its field dependence at
constant temperature. Figures 4 and 5 in Ref. \cite{Petrenko_99} give examples of such calculations.
The position of the specific heat maximum is not sensitive to the introduction of the relatively weak
next-to-nearest exchange interactions (such as were quoted in Ref. \cite{Kinney_79}, $J_2\!=\!-0.003$~K
and $J_3\!=\!0.010$~K), neither does it show any visible size-dependence. In the field dependence
of the specific heat, only one anomaly corresponding to the upper transition field is well-pronounced,
while there is no obvious anomaly corresponding to the lower transition field, which agrees with
previous MC simulations \cite{Schiffer_94}.  

\begin{figure}
\centerline{\psfig{figure=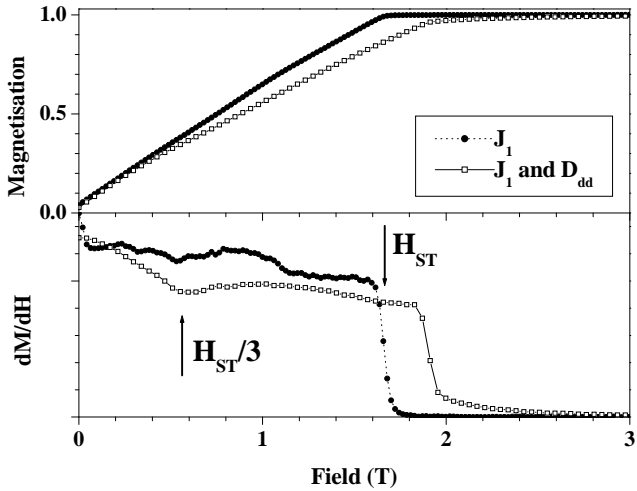,width=\columnwidth}}
\caption{Field dependence of the magnetisation (top) and susceptibility (bottom) at $T\!=\!2$~mK, 
$H\parallel (001)$. A $5\!\times\!5\!\times\!5$ model has been used to generate these data. Open and solid symbols represent the data for the model which included nearest neighbour exchange interaction with and without dipolar forces respectively.}
\label{M_H}
\end{figure}

In order to reproduce accurately the experimentally observed phase diagram of GGG, the simulation model must include nearest neighbour exchange interactions and also dipolar forces. However, even in the model including only nearest neighbour exchange more accurate calculations revealed signs of the phase transitions in a magnetic field. Fig.~\ref{M_H} presents the magnetisation curves and also their derivatives at $T\!=\!1$~mK for such models. Before reaching a saturation point at $H\!\approx\!1.7$~T, the raw magnetisation shows a relatively small change of the slope around $H\!\approx\!0.6$~T, which is not really a striking feature and therefore has passed unnoticed in our earlier calculations. In the susceptibility curves, however, a clear minimum is present at $H\!\approx\!0.6$~T. We believe that this minimum in susceptibility corresponds to the appearance of a collinear long-range ordered state induced by an applied magnetic field. In complete agreement with the theory~\cite{Zhit_2DAFM}, which analyses an order by disorder mechanism in various highly frustrated antiferromagnets, an ordering happens only around a special value of the magnetic field - one third of the saturation field in case of a garnet lattice.

\begin{figure}
\centerline{\psfig{figure=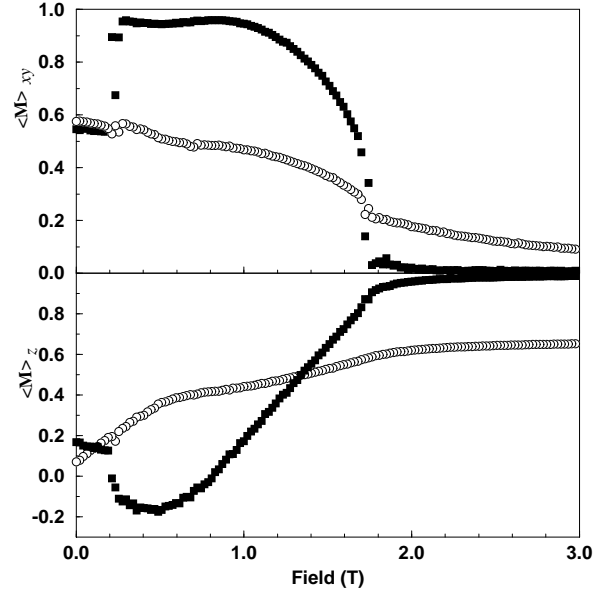,width=\columnwidth}}
\caption{Field dependence of parallel (bottom) and perpendicular (top) component of local magnetisation at $T$=0.1 K, $H\parallel (001)$. For notation see main text.} 
\label{OrPar_H}
\end{figure}

\begin{figure}
\centerline{\psfig{figure=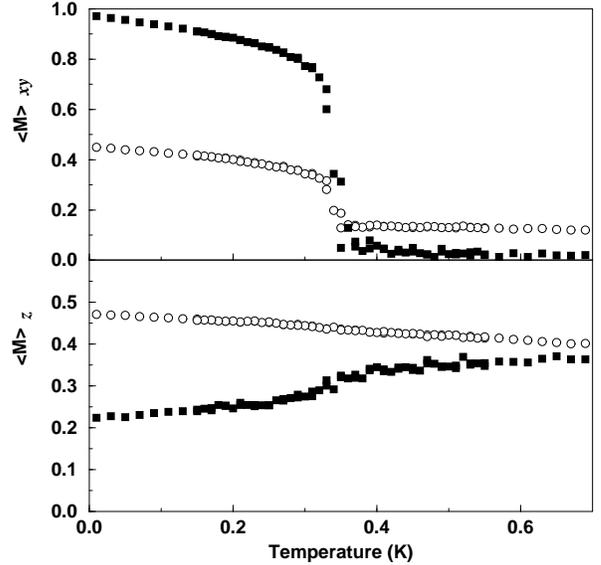,width=\columnwidth}}
\caption{Temperature dependence of parallel (bottom) and perpendicular (top) component of local magnetisation in a field $H$=1.06 T, $H\parallel (001)$. For notation see main text.} 
\label{OrPar_T}
\end{figure}

In GGG, that is in a model which includes the dipolar forces, an ordered magnetic structure induced by an
applied field is characterised by the appearance of a nonzero
average value of the perpendicular component of local magnetisation. The field dependence of parallel and 
perpendicular components of local magnetisation at constant temperature is shown on Fig. \ref{OrPar_H},
while Fig. \ref{OrPar_T} shows its temperature dependence in constant field. The two sets of
curves on each of these figures reflect the fact that in applied magnetic field the 24 Gd
sites are split unequally into two different symmetry sites -- ``A'' and ``B'' sites in the notation of
Ref. \cite{Schiffer_94}. When the field is applied along the $(001)$ direction, the 8 ``A'' sites (represented
by solid symbols on Fig. \ref{OrPar_H} and \ref{OrPar_T}) are of higher symmetry than the 16 ``B'' sites
(represented by open symbols). Clearly, $<M>_{xy}$ on the ``A'' sites serves as on order parameter
for the transition from a paramagnetic state into an \afmc ally ordered state. Spins on the ``B'' sites, however,
retain a nonzero value of the perpendicular component of magnetisation even in the paramagnetic state.
This effect is caused by the dipole-dipole interaction.

In order to avoid problems with possibly many metastable states the calculations were always started
at high temperatures and fields and then the system annealed as it came into equilibrium at the desired
field and temperature for measurement. However, even taking these precautions the problem of long
equilibration times at low-temperature low-field region was unavoidable. Therefore an abrupt jump 
of magnetisation around $H=0.25$~T clearly visible on Fig. \ref{OrPar_H} is most likely to be an 
artificial result.

Even without detailed knowledge of the magnetic structure in a field we can check how stable it is to 
the introduction of second and third next to nearest exchange interactions, $J_2$ and $J_3$. This has been
done by calculating the neutron scattering function $S_p(Q)$ for a $3\!\times\!3\!\times\!3$ system according
to formula (\ref{Sp}). The results suggest that the magnetic order is rather stable in all four quadrants in the
$J_2-J_3$ plane. The structure does not change when $J_2\!=\!-0.003$~K and $J_3\!=\!0.010$~K are
introduced corresponding to the values quoted in Ref. \cite{Kinney_79}.

In an applied magnetic field, where LRO is developed, the formula (\ref{Sp}) is no longer valid.  
Although it unambiguously shows the appearance of magnetic Bragg peaks, their intensity is not
calculated correctly.  However the overall field dependence of the intensity mimics extremely 
well the experimental data \cite{Petrenko_99}. There are two different groups of magnetic peaks, 
ferromagnetic and \afmc. The intensity of the former group is growing in lower fields and 
saturating in a higher field. The intensity of the latter group also grows in lower fields reaching a
maximum at around $H=1$~T and then decreases in higher fields and disappears above $H=2$~T. Exactly
the same behaviour has been seen by simulating single-crystal scattering intensity according to
formula~(\ref{Sxt}).  Bottom panel of Fig. \ref{Sq} shows the results of such calculations for $H=1.06$~T and
$T=200$~mK. In  an applied magnetic field a set of strong and sharp Bragg peaks replaces diffuse magnetic
scattering observed in zero field. One interesting aspect of the calculations must be emphasised here:
the relative intensity of the symmetry related \afmc\ peaks, such as, for example, $(210)$ and $(120)$, is not
constant in time. The intensity of each peak may change arbitrarily at any time from almost zero up to
maximum value, while the sum of two peaks intensities remains constant. 

The only discrepancy between the \MC\ results and the neutron scattering data in magnetic field 
is the presence in the later of an incommensurate peak located between two \afmc\ peaks, 
$(200)$ and $(210)$. While this incommensurate peak is clearly visible in the neutron scattering 
data~\cite{Petrenko_99}, neither $S_p(Q)$ nor $S_{xt}(Q)$ (including a model where the magnetic field is 
applied along the $(110)$ and $(111)$ directions) demonstrates peaks at an incommensurate position.
The reason for the discrepancy remains unknown at the moment.

\section{Conclusions} 

To summarise, we have presented the results of classical \MCS s for the low temperature behaviour of the
frustrated \afm\ on a garnet lattice. We have studied several different models, paying particular attention to two
of them. The first model, which includes only nearest-neighbour exchange interactions, does not order down to
lowest temperature, neither does it show any signs of spin-freezing. Calculations of the zero-temperature
specific heat for such model suggest the presence of soft modes. The indications of the phase transition into
an ordered (presumably collinear) state have been found at low temperature in applied magnetic field around a
third of the saturation field. The experimentally measured properties of GGG including the {\it H vs. T} magnetic
phase diagram
are consistent with our findings for the simulation model which includes nearest neighbour exchange interactions
and also dipolar forces. A perpendicular component of local magnetisation serves as an order parameter for the
phase transition in an applied field.

In conclusion we discuss several questions, which have been considered in this article, but which most certainly
require further theoretical investigations. 

Firstly, both the low-temperature specific heat, $C(\!T=\!0)$, and the single crystal scattering
function, $S_{xt}(Q)$, are unusually sensitive to the boundary conditions. The influence of particular boundary
conditions on the appearance of soft modes and incommensurate peaks in $S_{xt}(Q)$ needs to be examined further.

Secondly, in the ordered state only the total intensity for the pairs of a symmetry related
\afmc\ Bragg peaks, remains constant, while the intensity of an individual peaks may change.
Could this behaviour be related to the energy-free motion of long chains of magnetic moments, 
similar to what happens in the \Kag\ lattice, or is it a sign of the domain walls movement?

We are grateful to J.~T.~Chalker, M.~J.~P.~Gingras, M.~J.~Harris, C.~L.~Henley, P.~C.~Holdsworth, R.~Moessner
and M.~E.~Zhitomirsky for discussions and other valuable contributions and also to J.~N.~Reimers for his help
with the computer calculations on the initial stages of the project.

\end{document}